\documentclass[aip,jap,reprint,a4paper,floatfix,hidelinks]{revtex4-1}
\pdfoutput=1
\usepackage{graphicx}%
\usepackage{dcolumn}%
\usepackage{bm}%
\usepackage{amsmath}
\usepackage{siunitx}
\usepackage{float}
\usepackage{subfig}

\begin{document}
\title{Nanointerferometric Amplitude and Phase Reconstruction of Tightly Focused Vector Beams}

\author{Thomas Bauer}
\email{thomas.bauer@mpl.mpg.de}
\affiliation{Max Planck Institute for the Science of Light, Guenther-Scharowsky-Str. 1/Bldg. 24, 91058 Erlangen, Germany}
\affiliation{Institute of Optics, Information and Photonics, University Erlangen-Nuremberg, Staudtstr. 7/B2,  91058 Erlangen, Germany}

\author{Sergej Orlov}
\affiliation{Max Planck Institute for the Science of Light, Guenther-Scharowsky-Str. 1/Bldg. 24, 91058 Erlangen, Germany}
\affiliation{Institute of Optics, Information and Photonics, University Erlangen-Nuremberg, Staudtstr. 7/B2,  91058 Erlangen, Germany}

\author{Ulf Peschel}
\affiliation{Institute of Optics, Information and Photonics, University Erlangen-Nuremberg, Staudtstr. 7/B2,  91058 Erlangen, Germany}
\affiliation{Cluster of Excellence "Engineering of Advanced Materials", Naegelsbachstr. 49b, 91052 Erlangen, Germany}

\author{Peter Banzer}
\affiliation{Max Planck Institute for the Science of Light, Guenther-Scharowsky-Str. 1/Bldg. 24, 91058 Erlangen, Germany}
\affiliation{Institute of Optics, Information and Photonics, University Erlangen-Nuremberg, Staudtstr. 7/B2,  91058 Erlangen, Germany}

\author{Gerd Leuchs}
\affiliation{Max Planck Institute for the Science of Light, Guenther-Scharowsky-Str. 1/Bldg. 24, 91058 Erlangen, Germany}
\affiliation{Institute of Optics, Information and Photonics, University Erlangen-Nuremberg, Staudtstr. 7/B2,  91058 Erlangen, Germany}

\date{\today}

\begin{abstract}

\end{abstract}

\maketitle

\textbf{Highly confined vectorial electromagnetic field distributions represent an excellent tool for detailed studies in nano-optics and high resolution microscopy, such as nonlinear microscopy\cite{Masihzadeh2009}, advanced fluorescence imaging\cite{Hell1994,Maurer2010,Mudry2012} or nanoplasmonics\cite{Failla2006,Banzer2010}. Such field distributions can be generated, for instance, by tight focussing of polarized light beams\cite{RichardsWolf1959,Quabis2000,Youngworth2000,Dorn2003}. To guarantee high quality and resolution in the investigation of objects with sub-wavelength dimensions, the precise knowledge of the spatial distribution of the exciting vectorial field is of utmost importance. Full-field reconstruction methods presented so far involved, for instance, complex near-field techniques\cite{Grosjean2010,Lee2007}. Here, we demonstrate a simple and straight-forward to implement measurement scheme and reconstruction algorithm based on the scattering signal of a single spherical nanoparticle as a field-probe. We are able to reconstruct the amplitudes of the individual focal field components as well as their relative phase distributions with sub-wavelength resolution from a single scan measurement without the need for polarization analysis of the scattered light. This scheme can help to improve modern microscopy and nanoscopy techniques.}

In the optical analysis of sub-wavelength objects such as cellular structures\cite{Stephens2003}, plasmonic particles\cite{Failla2006,Banzer2010,Rodriguez2010} or single spins\cite{Maurer2010}, nano-optical tools including highly resolving microscopy techniques are used. Because such methods utilise complex and highly confined vector fields, the exact knowledge of the corresponding spatial field distributions is crucial. In the last decades, several techniques have been proposed to map these focal fields, such as using metal knife edges\cite{Quabis2001,Marchenko2011} to probe the total electric energy density distribution, fluorescence molecules\cite{Novotny2001}, tapered fibres\cite{Rhodes2002,Bouhelier2003} or tip-based methods\cite{Lee2007} to image specific field orientations, or near-field scanning optical microscope (NSOM) techniques to extract amplitude and even phase information\cite{Grosjean2010}. These NSOM-based methods require complex measurement and detection schemes and calibration procedures to allow for amplitude and phase mapping of individual field components. As an alternative approach for measuring phase information also a single particle scattering scheme was proposed recently\cite{Garbin2009}, where the authors show that Mie-scattering can distinguish the topological charge of vortex-beams.

\begin{figure}[tbh]
\includegraphics[width=0.85\linewidth]{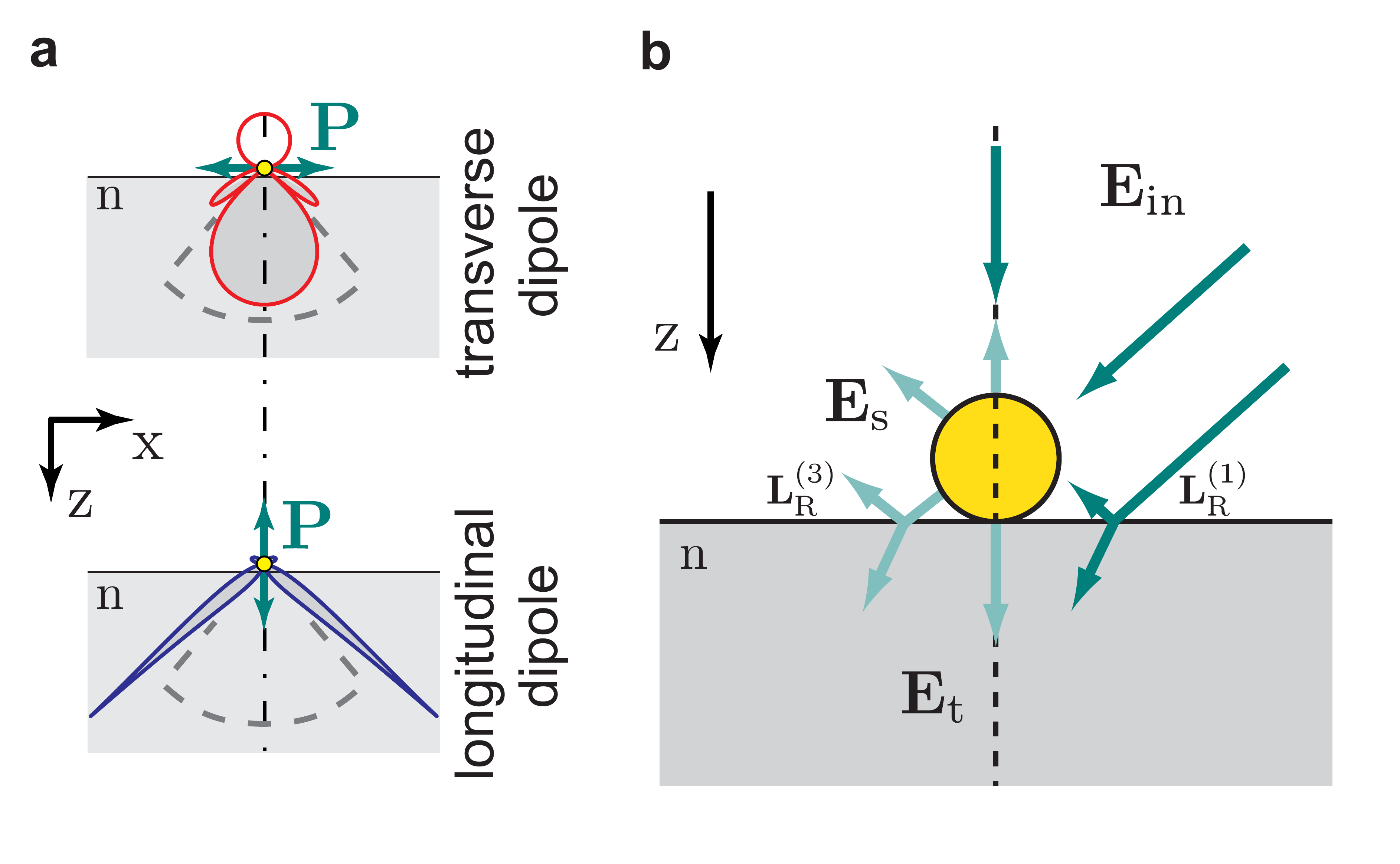}
\caption{Schematics of the scattering process. \textbf{a} Far-field emission pattern of differently oriented electric dipoles on an interface with refractive index $n=1.5$. The dipole intensity emitted into a limited solid angle (dashed lines) shows a strong dependence on the dipole direction. \textbf{b} The interface additionally leads to a coupling of the reflected field from the substrate $\mathbf{L}_R^{(1,3)}\cdot \mathbf{E}_\text{in/s}$ and the scattered field off the spherical scatterer $\mathbf{E}_\text{s}$.}
\label{fig:emissionpattern_theory}
\end{figure}

We now demonstrate a precise and easily implementable field reconstruction technique for highly confined field distributions created by arbitrary focusing systems, based on, what we call, Mie-scattering nanointerferometry. The basic concept of this reconstruction method for highly confined focal field distributions can be understood as follows. We use a metallic nanosphere on a glass substrate as local field sensor and scan it step-wise through the focal field distribution under investigation. If the particle size is chosen small enough, it will be excited by the local electric field $\mathbf{E}(\mathbf{r}_0)$ only. Its response can be described, in a first approximation, by an electric dipole. Hence, for a particle on a substrate differently oriented electric dipole moments will be excited in the particle for different positions in the field distribution under investigation depending on the corresponding local field orientation, and resulting in different scattering patterns \cite{Courtois1996} (see Fig.\ref{fig:emissionpattern_theory}(a)). Here we show, that the initial local field and therefore the three-dimensional focal field distribution can be determined accurately in amplitudes and relative phases of the individual field components by collecting the transmitted light angularly resolved\cite{Jasny1997} and exploiting the interference signal between incoming and scattered field. The interference and, hence, the phase information is preserved herein by effectively changing the observation direction.

To analytically describe the full scattering process, the unknown focal field is expanded into electromagnetic multipoles. With this choice of a basis system, only the lower expansion orders have to be taken into account for highly confined fields (see supplementary section S.6 and \onlinecite{Hoang2012,Mojarad2008}). The focal electric field distribution can thus be expressed as
\begin{equation}
\mathbf{E}_\mathrm{in}(\mathbf{r}) = \sum_{n=1}^{\infty}\sum_{m=-n}^{n} A_{mn}\mathbf{N}_{mn}(\mathbf{r}) + B_{mn} \mathbf{M}_{mn}(\mathbf{r}),
\end{equation}
where $\mathbf{N}_{mn}$ and $\mathbf{M}_{mn}$ are regular vector spherical harmonics representing differently oriented electric and magnetic multipoles\cite{Tsang2000,Orlov2012} expanded around the geometrical focus. This transfers the full information of the unknown focal field from the complex-valued electric field components $\mathrm{E}_\text{in,i}(\mathbf{r})$ at each point in the focal plane to the complex-valued multipole expansion coefficients $A_{mn}$ and $B_{mn}$. This representation allows to relate distinct far-field patterns with the focal field distribution under investigation, where electric dipoles can be associated with the local electric field $\mathbf{E}_\text{in}(\mathbf{r}_0)$ at the expansion point $\mathbf{r}_0$, electric quadrupoles and magnetic dipoles with its gradient, and higher order multipoles with higher order moments of the field. In addition, it permits a simple treatment of scattering problems via the T-matrix approach\cite{Mishchenko2002}.

The focal electric field distribution can thus either be reconstructed by determining all multipole coefficients of the focal field at one single point or by scanning the dipole contributions in the focal plane and relating them to one common expansion point via the translation theorem for vector spherical harmonics\cite{Cruzan1962}. To achieve the in both cases necessary high position accuracy and repeatability, the probe is  immobilized on a substrate in our experiments, which introduces an interface in the scattering problem (see Fig. \ref{fig:emissionpattern_theory}). The resulting transmitted field $\mathbf{E}_\mathrm{t}$ can then be expressed as the incoming field $\mathbf{E}_\mathrm{in,t}$ and a scattering field term $\mathbf{E}_\mathrm{s,t}$, incorporating the full geometry of the system:
\begin{align}
\mathbf{E}_\mathrm{t}(\mathbf{r}) = \mathbf{E}_\mathrm{in,t}(\mathbf{r}) + \mathbf{E}_\mathrm{s,t}(\mathbf{r}).
\end{align}
For a simplification of the description of the interface, we transform the basis functions in the forward hemisphere to $\mathbf{N}_{mn,t}$ and $\mathbf{M}_{mn,t}$, leaving the expansion coefficients $A_{mn}$ and $B_{mn}$ unaltered and thus keeping the low order multipole expansion (see supplementary section S.1). The scattered electric field $\mathbf{E}_\text{s,t}$ is then related to the incoming field $\mathbf{E}_\text{in,t}$ by describing the interaction of the nanoprobe and the substrate with the incoming light field by an effective scattering matrix $\mathbf{T}_\mathrm{eff}$, depending only on the known geometry and refractive indices as:
\begin{align}
\mathbf{E}_\mathrm{s,t} &= \left(1-\mathbf{T} \mathbf{L}_\mathrm{R}^{(3)}\right)^{-1} \mathbf{T} \left(1+\mathbf{L}_\mathrm{R}^{(1)}\right)  \mathbf{E}_\mathrm{in,t}  = \mathbf{T}_\mathrm{eff} \mathbf{E}_\mathrm{in,t},
\end{align}
with $\mathbf{T}$ being the scattering matrix of the probe particle (including information about the particle shape, size and its optical properties) and $\mathbf{L}_R^{(1,3)}$  being reflection operators of the substrate (see Fig.\ref{fig:emissionpattern_theory}(b) and sup\-ple\-men\-ta\-ry section S.1).

\begin{figure}[tbh]
\includegraphics[width=0.9\linewidth]{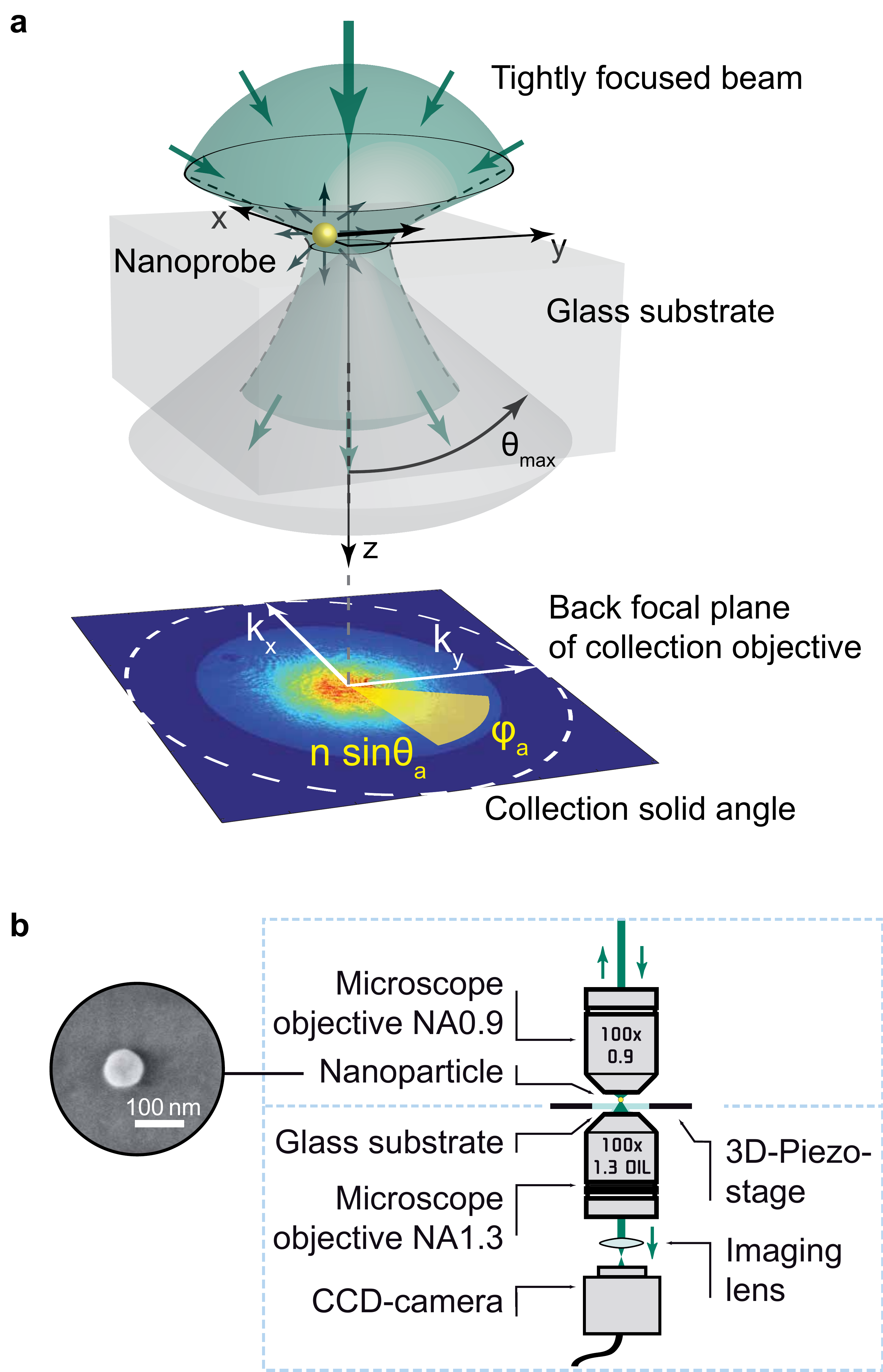}
\caption{Sketch of the experimental implementation of the reconstruction scheme. \textbf{a} A spherical metallic nanoparticle adhered on a glass substrate is scanned through the focal field distribution under investigation. The scattered and transmitted intensity is consequently collected with a variable solid angle $(\theta_a, \varphi_a)$. \textbf{b} Experimental setup: an arbitrarily structured input beam is focused by a high numerical aperture microscope objective (NA = 0.9). The nanoprobe can be precisely scanned through the focal plane by means of a 3D-piezo stage. The transmitted light is then collected by an immersion-type microscope objective (NA = 1.3) and its back-focal-plane is imaged onto a CCD-chip. The reflected light can be measured in addition to ensure the exact focus position.}
\label{fig:setup}
\end{figure}

\begin{figure*}[tbh]
\includegraphics[width=\linewidth]{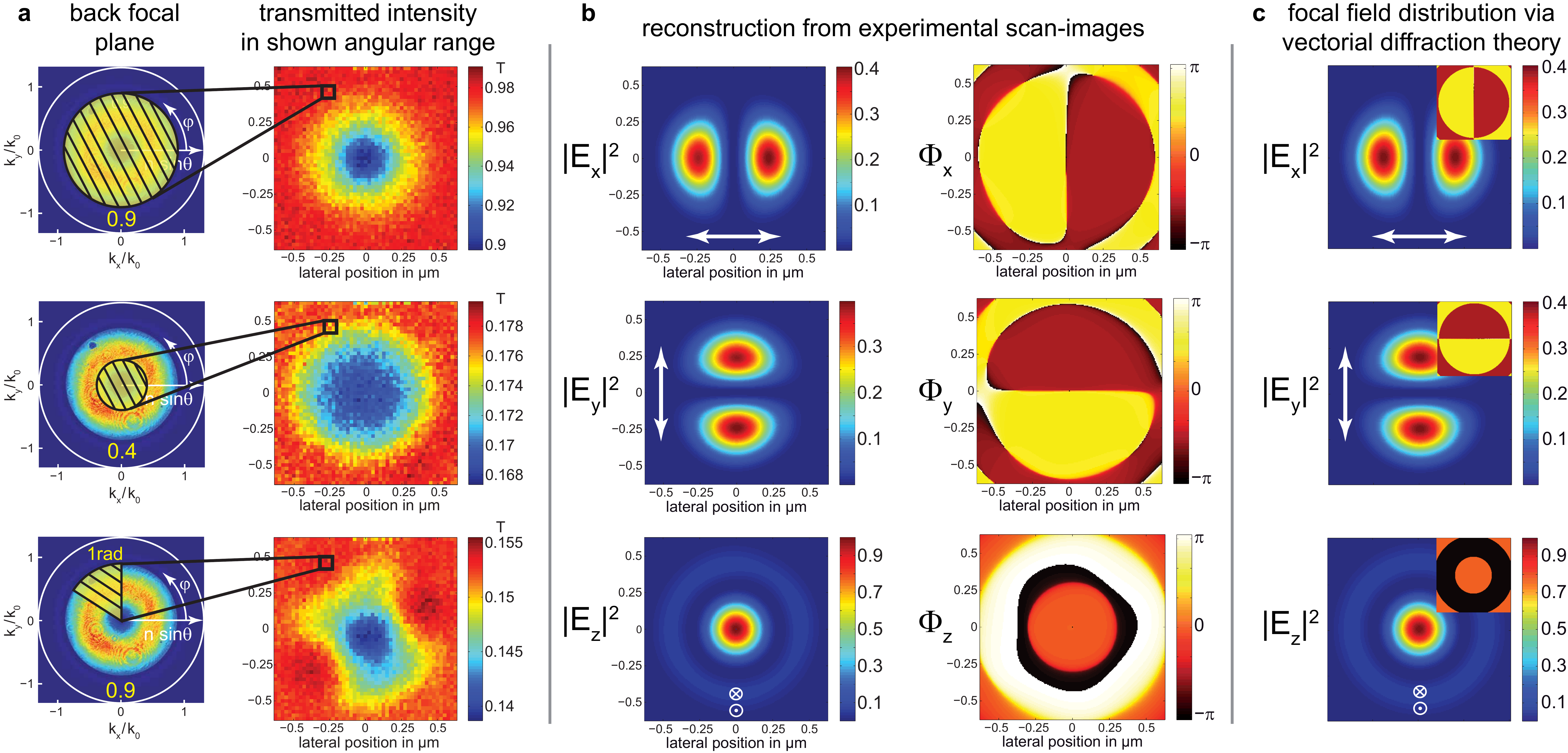}
\caption{Experimental results and theoretical comparison for a radially polarized vector beam. \textbf{a} Image of the back focal plane of the collection objective with an NA of 1.3 for one position of the nanoprobe relative to the focal field. The measured intensity scan-images in transmission for a wavelength of \SI{530}{\nano\metre} correspond to three different collecting solid angles with a full aperture of NA  0.9 and 0.4 as well as an azimuth angle of $\varphi_a=1\ \text{rad}$ for an NA of 0.9,  all derived from the same measured back focal plane images for different probe positions. The intensity is normalized to the total intensity of the input beam. \textbf{b} Squared electric field components $\left| E_\text{i} \right|^2$ and relative phases $\Phi_\text{i}$ in the focal plane reconstructed from the measured intensity distributions up to a multipole order of $n=8$. \textbf{c} Energy density distribution for the three field components in the focal plane of the same beam calculated via vectorial diffraction theory. The insets show the calculated phase distribution in the same scale and colormap as in (b).}
\label{fig:results}
\end{figure*}

To relate the unknown input field to an experimentally measurable quantity, the resulting power transmitted at an angle $(\theta,\varphi)$ (see Fig. \ref{fig:setup}(a)) can then be expressed by
\begin{equation}
\text{P}_\mathrm{t}(\theta,\varphi) = \text{P}_\text{in}(\theta,\varphi) + \text{P}_\text{s}(\theta,\varphi) + \text{P}_\text{ext}(\theta,\varphi),
\end{equation}
following the classical Mie scattering problem\cite{Stratton1940}with 
\begin{align*}
\text{P}_\text{in}&=\frac{1}{2}\mathrm{Re} \left[\mathbf{E}^*_\mathrm{in,t} \times \mathbf{H}_\mathrm{in,t}\right], \quad \text{P}_\text{s}=\frac{1}{2}\mathrm{Re} \left[\mathbf{E}_\mathrm{s,t}^* \times \mathbf{H}_\mathrm{s,t}\right], \\
\text{P}_\text{ext}&=\frac{1}{2}\mathrm{Re} \left[\mathbf{E}_\mathrm{in,t}^* \times \mathbf{H}_\mathrm{s,t} + \mathbf{E}_\mathrm{s,t}^* \times \mathbf{H}_\mathrm{in,t}\right]
\end{align*}
as the incoming, scattered and extinct power measured in transmission.
Here, the magnetic field components are determined from the electric field components in the far-field using the plane wave spectrum of the transmitted multipoles (see supplementary section S.1).
The interference term $\text{P}_\text{ext}$, which depends on both $\theta$ and $\varphi$, not only allows for the extraction of amplitude information of the multipole expansion coefficients, but also the phase relation between them.
It can be shown, that for an unambiguous reconstruction of both, amplitudes and relative phases of the individual electric field components in the focal plane the combination of the following collection schemes is sufficient. First, the transmitted power $\text{P}_\text{t}$ has to be recorded and integrated in a given solid angle around the optical axis for each position of the particle in the focal plane (see Fig. \ref{fig:results}(a) top and center). For that purpose, an integration in $\theta$ from $0$ to $\theta_{a}$ and in $\varphi$ from $0$ to $2\pi$ is performed. By this integration over the whole range in $\varphi$, some phase information is lost. Thus, to regain the full information about the relative phases, an effective break of the cylindrical symmetry of the collection system (objective) is introduced. This can be realized by secondly choosing a collection sector, hence performing a $\theta$-integration from $0$ to $\theta_{a}$ and a $\varphi$ integration from $\varphi_{1}$ to $\varphi_{2}$ (with $\vert\varphi_{1} - \varphi_{2}\vert < 2\pi$; see Fig. \ref{fig:results}(a) bottom). With this choice, the light emitted into a certain direction relative to the optical axis is analysed. Both schemes can be realised easily in the experiment by imaging the transmitted power angularly resolved in the back focal plane of a high numerical aperture (NA) collection lens for each position of the nanoprobe relative to the beam. The integration ranges can then be chosen freely, with $\theta$ bound by the maximum collection angle $\theta_\text{max}$ of the employed collection lens.
Additionally collected angular ranges can furthermore reduce the influence of noise in the system (see supplementary section S.2).
With the known scattering matrix $\mathbf{T}_\mathrm{eff}$, the powers transmitted into different angular ranges for each position of the nanoprobe relative to the input beam thus establish a system of positive quadratic forms of the unknown multipole expansion coefficients. The inversion of this equation system then leads to the vectorial focal electric field distribution.

For the experimental demonstration of the introduced nanointerferometric vectorial field reconstruction scheme a custom-made scanning setup (similar to Ref. 11) is used, which represents a scattering system  that is easily applicable to different research areas. The highly confined focal field distribution under investigation is created by sending a collimated input beam with a wavelength of 530 nm into a high NA objective. The probe system which we use for scanning consists of a spherical gold nanoparticle with a diameter of $\SI{82}{\nano\metre}$ on a glass substrate (see Fig. \ref{fig:setup}). The diameter of the particle was chosen such, that it is small enough to exhibit predominately a dipolar response but still large enough to guarantee a large scattering cross-section for a high signal to noise ratio. Furthermore, the spherical symmetry of the particle simplifies the scattering matrix $\mathbf{T}$. The particle is scanned through the focal field distribution and at each point the light transmitted through the substrate and scattered off the nanoprobe in the forward direction is collected by an immersion-type microscope objective (see Fig. \ref{fig:setup}(b)). The back focal plane of the collection objective is then imaged onto a CCD-camera, providing access to the k-spectrum of the transmitted light. Integrating this angularly resolved transmitted intensity over different angular ranges $(\theta_a, \varphi_a)$ (see Fig. \ref{fig:setup}(a)), enables the reconstruction of the vectorial focal field distribution as discussed above. The collection objective is hereby assumed to only introduce negligible additional errors to the collected intensity distribution, which can be achieved with modern high quality oil-immersion objectives. It is worth noting here, that the aforementioned scheme is also applicable to reflection-type setups and probes embedded in a homogeneous medium. 

By choosing a radially polarized doughnut beam as exemplary input field, a three-dimensional vectorial field distribution is generated under tight focusing conditions, exhibiting an on-axis longitudinal field component\cite{Dorn2003,Youngworth2000} and off-axis transverse field components (see Fig. \ref{fig:results}(c), calculated via vectorial diffraction theory\cite{RichardsWolf1959}). Such a tightly focused cylindrical vector beam finds applications in several fields of nano-optics and imaging\cite{Volpe2009,Banzer2010_2}.

In Fig. \ref{fig:results}(a), the experimental results are shown. For every position of the nanoparticle (scan step-size: $\SI{25}{nm}$; see supplementary section S.5) a single image of the back focal plane is recorded (see Fig. \ref{fig:results}(a) left column for one position of the particle and supplementary section S.3). From this camera data, two-dimensional scan images are derived by plotting the transmitted intensity integrated over the corresponding angular range in the back focal plane image for each probe position (see Fig. \ref{fig:results}(a) right column). Here, we choose three different angular ranges for integration in the measured back focal plane images to reduce the influence of experimental noise (see supplementary section S.2). We choose two solid angles which correspond to two full numerical apertures of 0.9 and 0.4 and a sector of $\varphi_a = 1\, \text{rad}$ at an NA of 0.9.
The non-rotational symmetric collection angle preserves the interference information (see Fig. \ref{fig:results}(a) bottom). This signal is highly sensitive to the actual optical properties of the nanoprobe, which were determined experimentally for the selected nanoprobe and wavelength to $\epsilon_\text{AuNP} = -3.0+2.1 \imath$ (see supplementary section S.4).
The electric energy densities and phases of the focal field components reconstructed from these intensity distributions (see Fig. \ref{fig:results}(b)) show a very good overlap with the calculated field components shown in Fig. \ref{fig:results}(c). Small deviations from theory are visible in the profiles reconstructed form the experimental data. We attribute these differences to small aberrations of the focussing system. Thus, deviations from the ideally expected focal field distribution can be reconstructed with sub-wavelength precision.

In summary, we have shown an easily applicable reconstruction scheme to determine the full vectorial amplitude and relative phase distributions of highly confined electromagnetic fields. The technique relies on the nanointerferometry between the input field and the field scattered off a nanoprobe as well as an angularly resolved measurement of the resulting far-field intensity. By adapting the described scheme, also  a reconstruction of field distributions for other focusing systems can be realised, including the near-field distribution of NSOM-tips. 

\section*{Methods}
\subsection*{Experimental Setup}\vspace*{-1em}
A tunable light source (NKTphotonics SuperK PowerPlus with an AA Opto-Electronic MDSnC-TN acousto-optical tunable filter) is providing a linearly polarized Gaussian beam at a wavelength of $\SI{530}{\nano\metre}$, which is optionally converted into a radially or azimuthally polarized doughnut beam by a liquid crystal radial polarization converter (ARCoptix). The beam is then guided via four mirrors top-down onto a microscope objective with a NA of $0.9$ (Leica HCX PL FL 100x/0.90 POL 0/D). The beam radius ($1/e^2$) of the doughnut beam is \SI{1.64}{\milli\metre}, while the diameter of the entrance aperture of the microscope objective is \SI{3.6}{\milli\metre}. A single spherical gold nanoparticle with a diameter of \SI{82}{\nano\metre} sitting on a glass cover slip is scanned through the focal plane by a high precision 3D piezo table (PI P-527) and the transmitted light is collected via an oil immersion objective with a NA of 1.3 (Leica HCX PL FLUOTAR 100x/1.30 OIL). In addition, the reflected light is collected by the focusing objective. The  angular distributions of the transmitted and optionally the reflected light are detected via imaging the back focal plane of the corresponding objectives on CCD-cameras (The Imaging Source DMK 23G618).

\subsection*{Sample preparation}\vspace*{-1em}
The spherical gold nanoparticle was fabricated at the Max Planck Institute for Polymer Research, Mainz, Germany by laser induced melting of a commercial colloidal gold solution. The resulting solution of spherical nanoparticles was drop-coated on a microscope cover slip with a thickness of \SI{170}{\micro\metre} and prestructured gold markers to reproducibly measure one single nanoparticle.

\begin{acknowledgments}
\vspace*{-1em}We gratefully acknowledge fruitful discussions with Martin Neugebauer and Sascha Batz, and thank Markus Schmelzeisen from the Max Planck Institute for Polymer Research in Mainz for the fabrication of the used scattering particle.
\end{acknowledgments}

\newpage
\setcounter{figure}{0}
\renewcommand{\thesubsection}{\normalsize S.\arabic{subsection}}
\renewcommand{\figurename}{Figure S\hspace*{-0.5ex}}
\section*{Supplementary materials}
This supplementary material describes in more detail the developed reconstruction algorithm. In addition, we show the characterisation results of the employed nanoprobe as well as the reconstruction of a tightly focused azimuthally polarised doughnut beam. Furthermore, the theoretical and experimental sources of errors are discussed.

\subsection{Detailed description of the influence of the substrate}
Describing the scattering process off the nanoprobe in the basis of vector spherical harmonics (VSHs), we follow the notation of Tsang et al.\cite{Tsang2000_S}. To simplify the theoretical description of the energy transfer through the interface and the interaction of the scattered fields with the latter, we transfer the effect of the interface to the basis functions $\mathbf{M}^{(j)} _{mn}$ and $\mathbf{N}^{(j)} _{mn}$, leaving the expansion coefficients of the initial focal field unaltered. For this purposes we need a plane wave representation of the first and third type VSHs, which results in \cite{Tsang2000_S}
\begin{align}
\mathbf{M}^{(j)} _{mn} (\mathbf{r}) &= \frac{\left(-\mathrm{i} \right)^n \gamma _{mn}}{2\pi\left(1 + \delta_{j,1} \right)}\int _{\Omega_j} \mathrm{d}\Omega_j \mathrm{e}^{\mathrm{i}\mathbf{k} \cdot \mathbf{r}} \mathbf{X}^{(2)}_{mn}\left(\theta_k, \varphi _k \right),   \nonumber \\
\mathbf{N}^{(j)} _{mn} (\mathbf{r}) &= \frac{\left(-\mathrm{i} \right)^{n-1} \gamma _{mn}}{2\pi\left(1 + \delta_{j,1} \right)} \int _{\Omega_j} \mathrm{d}\Omega_j \mathrm{e}^{\mathrm{i}\mathbf{k} \cdot \mathbf{r}} \mathbf{X}^{(1)}_{mn}\left(\theta_k, \varphi _k \right), 
\label{eq:MN_plane}
\end{align}
where $\gamma_{mn}$ are multipole dependent prefactors, $\mathbf{k}$ is the wave vector of the respective plane wave and $\mathbf{X}^{(1)}_{mn}$ and $\mathbf{X}^{(2)}_{mn}$ correspond to VSHs defined analogue to Ref. \onlinecite{Tsang2000_S}.
The index $j$ distinguishes the incoming and scattered fields, using VSHs of first and third type. The integration region of these two types differs due to the inclusion of evanescent field terms in the scattered fields and is $\Omega_1 =\varphi \in \left(0, 2\pi \right), \theta \in \left(0, \pi \right)$, and $\Omega_3 =\varphi \in \left(0, 2\pi \right), \theta \in \left(0, \pi/2 - \mathrm{i} \infty \right)$ if an observation point is in the upper hemisphere or $\Omega_3 =\varphi \in \left(0, 2\pi \right), \theta \in \left(\pi/2 + \mathrm{i} \infty, \pi \right)$ if it is in the lower hemisphere.

The effect of the interface can now be introduced by applying Fresnel equations to each plane wave component in Eq. (\ref{eq:MN_plane}). We here use the Fresnel transmission $t_{\eta}\left(\theta \right)$ and reflection $r_{\eta}\left(\theta \right)$ coefficients from Ref. \onlinecite{JDJack_S} (where $\eta=1$ for $\theta$ or p-polarised components and $\eta = 2$ for $\varphi$ or s-polarised components) . As a result of the reflection the basis vectors  assume the form
\begin{align}
\mathbf{M}^{(j)} _{mn,r} = &\frac{\left(-\mathrm{i} \right)^n \gamma _{mn}}{2\pi\left(1 + \delta_{j,1} \right)}\int _{\Omega_j} \mathrm{d}\Omega_j \mathrm{e}^{\mathrm{i}\mathbf{k}_R \cdot \mathbf{r}}\mathrm{e}^{2\mathrm{i}n_1k R_{sp} \cos \theta _k} \nonumber \\ \times & \sum _{\eta} r_{\eta}\mathbf{e}_{R\eta}\left[\mathbf{e}_{\eta}\cdot\mathbf{X}^{(2)}_{mn}\left(\theta_k, \varphi _k \right) \right],   \nonumber \\
\mathbf{N}^{(j)} _{mn,r} = & \frac{\left(-\mathrm{i} \right)^{n-1} \gamma _{mn}}{2\pi\left(1 + \delta_{j,1} \right)} \int _{\Omega_j} \mathrm{d}\Omega_j \mathrm{e}^{\mathrm{i}\mathbf{k}_R \cdot \mathbf{r}}\mathrm{e}^{2\mathrm{i}n_1k R_{sp} \cos \theta _k} \nonumber \\ \times & \sum _{\eta} r_{\eta}\mathbf{e}_{R\eta}\left[\mathbf{e}_{\eta}\cdot\mathbf{X}^{(1)}_{mn}\left(\theta_k, \varphi _k \right) \right].
\label{eq:MNref_plane}
\end{align}
Here, $\mathbf{k}_R$ is the wave vector of a reflected plane wave, $\mathbf{e}_{R\eta}$ are unitary vectors of the reflected coordinate system and $R_{sp}$ is the radius of the nanoprobe. To account for the interaction of the reflected field with the nanoprobe above the interface, the reflected multipoles (\ref{eq:MNref_plane}) have to be reexpanded into the non-reflected ones. Thus we write
\begin{align}
\mathbf{M}^{(j)} _{mn,r} = \frac{1}{\left(1 + \delta_{j,1} \right)}\sum _{\nu = 1}^{\infty}\gamma _{m,n\nu}\left[u^{(11)}_{m,n\nu} \mathbf{M}^{(j)} _{m\nu}+ \mathrm{i}u^{(12)}_{m,n\nu} \mathbf{N}^{(j)} _{m\nu}\right],   \nonumber \\
\mathbf{N}^{(j)} _{mn,r} = \frac{-\mathrm{i} }{\left(1 + \delta_{j,1} \right)}\sum _{\nu = 1}^{\infty}\gamma _{m,n\nu}\left[u^{(21)}_{m,n\nu} \mathbf{M}^{(j)} _{m\nu}+ \mathrm{i}u^{(22)}_{m,n\nu} \mathbf{N}^{(j)} _{m\nu}\right], 
\label{eq:MNref_plane2}
\end{align}
where
\begin{align}
u^{(11)}_{m,n\nu} = &\int _{\Omega_j} \mathrm{d}\theta \sin \theta \mathrm{e}^{2\mathrm{i}n_1k R_{sp} \cos \theta _k}\nonumber \\ 
\times & \left[-r_{\theta} \frac{m^2}{\sin ^2 \theta}P^m_nP^m_{\nu} + r_{\varphi} \frac{\partial P^m_n}{\partial \theta}\frac{\partial P^m_{\nu}}{\partial \theta} \right] ,   \nonumber \\
 u^{(12)}_{m,n\nu} = &\int _{\Omega_j} \mathrm{d}\theta \sin \theta \mathrm{e}^{2\mathrm{i}n_1k R_{sp} \cos \theta _k}\nonumber \\
 \times & \left[-r_{\theta}\frac{\mathrm{i}m}{\sin  \theta}P^m_n\frac{  \partial P^m_{\nu}}{\partial \theta}+ r_{\varphi} \frac{\partial P^m_n}{\partial \theta}\frac{\mathrm{i}m}{\sin  \theta}P^m_{\nu} \right], \nonumber \\
u^{(21)}_{m,n\nu} = &\int _{\Omega_j} \mathrm{d}\theta \sin \theta \mathrm{e}^{2\mathrm{i}n_1k R_{sp} \cos \theta _k}\nonumber \\
\times & \left[r_{\theta}\frac{\mathrm{i}m}{\sin  \theta}P^m_{\nu}\frac{  \partial P^m_n}{\partial \theta}- r_{\varphi} \frac{\partial P^m_{\nu}}{\partial \theta}\frac{\mathrm{i}m}{\sin  \theta}P^m_n \right], \nonumber \\
u^{(22)}_{m,n\nu} = &\int _{\Omega_j} \mathrm{d}\theta \sin \theta \mathrm{e}^{2\mathrm{i}n_1k R_{sp} \cos \theta _k}\nonumber \\
\times & \left[-r_{\theta}\frac{\partial P^m_n}{\partial \theta}\frac{\partial P^m_{\nu}}{\partial \theta}+ r_{\varphi} \frac{m^2}{\sin ^2 \theta}P^m_nP^m_{\nu}\right],
\label{eq:MNref_plane3}
\end{align}
and 
\begin{equation}
\gamma _{m,n\nu} =\left(-1 \right)^{m+\nu} \frac{\mathrm{i} ^{\nu-n} \gamma _{mn}}{\gamma _{m\nu}}\frac{2\nu + 1}{\nu \left(\nu + 1 \right)}\frac{\left(\nu - m \right)!}{\left(\nu + m \right)!}.
\label{eq:gamma22}
\end{equation}
The elements of the reflection operators $\mathbf{L}_R^{(1,3)}$ can then be readily obtained from Eq. (\ref{eq:MNref_plane2}).

Concerning the initial and scattered fields transmitted through the interface, the free space multipoles can be similarly expressed as 
\begin{align}
\mathbf{M}^{(j)} _{mn,t} = &\frac{\left(-\mathrm{i} \right)^n \gamma _{mn}}{2\pi\left(1 + \delta_{j,1} \right)}\int _{\Omega_j} \mathrm{d}\Omega _j \mathrm{e}^{\mathrm{i}\mathbf{k}_T \cdot \mathbf{r}}\mathrm{e}^{\mathrm{i}n_1k R_{sp} \cos \theta _k} \nonumber \\ \times & \sum _{\eta} t_{\eta}\mathbf{e}_{T\eta}\left[\mathbf{e}_{\eta}\cdot\mathbf{X}^{(2)}_{mn}\left(\theta_k, \varphi _k \right) \right],   \nonumber \\
\mathbf{N}^{(j)} _{mn,t} = &\frac{\left(-\mathrm{i} \right)^{n-1} \gamma _{mn}}{2\pi\left(1 + \delta_{j,1} \right)} \int _{\Omega_j} \mathrm{d}\Omega _j \mathrm{e}^{\mathrm{i}\mathbf{k}_T \cdot \mathbf{r}}\mathrm{e}^{\mathrm{i}n_1k R_{sp} \cos \theta _k} \nonumber \\ \times & \sum _{\eta} t_{\eta}\mathbf{e}_{T\eta}\left[\mathbf{e}_{\eta}\cdot\mathbf{X}^{(1)}_{mn}\left(\theta_k, \varphi _k \right) \right], 
\label{eq:MNtra_plane}
\end{align}
Thus, when an electromagnetic field represented as a vector $\begin{pmatrix}B_{mn}\\A_{mn}\end{pmatrix}$ containing expansion coefficients of the free-space multipoles is transmitted through a planar interface, the electromagnetic multipoles are replaced by their transmitted (\ref{eq:MNtra_plane}) counterparts but the expansion coefficients of the transmitted electric field remain unchanged. This fact enables us to write the density of the Poynting vector in the hemisphere of the substrate in a matrix form as
\begin{align}
\mathbf{P}_\text{in} = \frac{1}{2}\mathrm{Re} \left[\mathbf{E}^{*}_\text{in} \mathbf{w}^{(T)}_{i} \mathbf{E}_\text{in}\right],\quad \mathbf{P}_\text{s} = \frac{1}{2}\mathrm{Re} \left[\mathbf{E}^{*}_\text{s}  \mathbf{w}^{(T)}_{s} \mathbf{E}_\text{s}\right], \nonumber \\
\mathbf{P}_\text{ext} = \frac{1}{2}\mathrm{Re}\left[\mathbf{E}^{*}_\text{in} \mathbf{w}^{(T)}_{e} \mathbf{E}_\text{s} + \mathbf{E}^{*}_\text{s} \mathbf{w}^{(T)}_{e} \mathbf{E}_\text{in} \right].
\label{eq:Poynt_subs3}
\end{align}
where the matrices $\mathbf{w}^{(T)}_{i}$, $\mathbf{w}^{(T)}_{s}$ and $\mathbf{w}^{(T)}_{e}$ contain scalar products of the plane wave spectrum functions of the transmitted multipoles. The energies transmitted to a solid angle $(\theta_a, \varphi_a)$ can then be obtained by performing an integration of those matrices over the desired solid angle and applying (\ref{eq:Poynt_subs3}).

\subsection{Mathematical background of the reconstruction algorithm}
From a mathematical point of view the density of the Poynting vector as represented in Eqs. (\ref{eq:Poynt_subs3}) is a positively defined quadratic form, which has a unique solution. However, when an integration is performed over a given solid angle, the rank of the matrix may decrease (for example, if $\varphi = (0, 2\pi)$). Basically this means, that for some combinations of $m$ and $n$ the crossproducts of the multipole amplitudes (like $A_{ij}A^{*}_{i'j'}$ or $A_{ij}B^{*}_{i'j'}$) do not influence the experimentally observed transmission. Therefore, in order to preserve the interferometric information, the integration region has to be chosen carefully, so that the rank of the matrix does not decrease. Then, we end up with a number of quadratic equations, which can be solved using a variety of different techniques.

The most straightforward inversion approach is based on a relinearisation of quadratic equations\cite{Golub1996_S}, where one introduces new variables $F_l=A_iA_j$ (here $A_i$ is either the real or imaginary part of the multipole amplitude), which are linearly related to the transmission at each scan position. In an ideal case, the unknowns $F_l$ are trivially retrieved and thus the unknown multipole amplitudes are obtained. However, in an experimental situation, one always has to account for some systematic and statistical perturbations, therefore a straightforward matrix inversion of  the form $\mathbf{A}\mathbf{x}-\mathbf{y} = \mathbf{\xi}$ may fail. Depending on the strength of the stochastic term $\mathbf{\xi}$ relative to the real signal, some quadratic forms may become singular in the sense, that some crossproducts for a given integration region influence the transmission signal less than it is influenced by the stochastic term. This can be counteracted by a careful choice of additional integration regions, where that crossproduct contributes stronger to the transmission.

\subsection{Sampling effect in the experimental CCD-images}
\begin{figure}[hbt]
\centering
\includegraphics[width=0.7\linewidth]{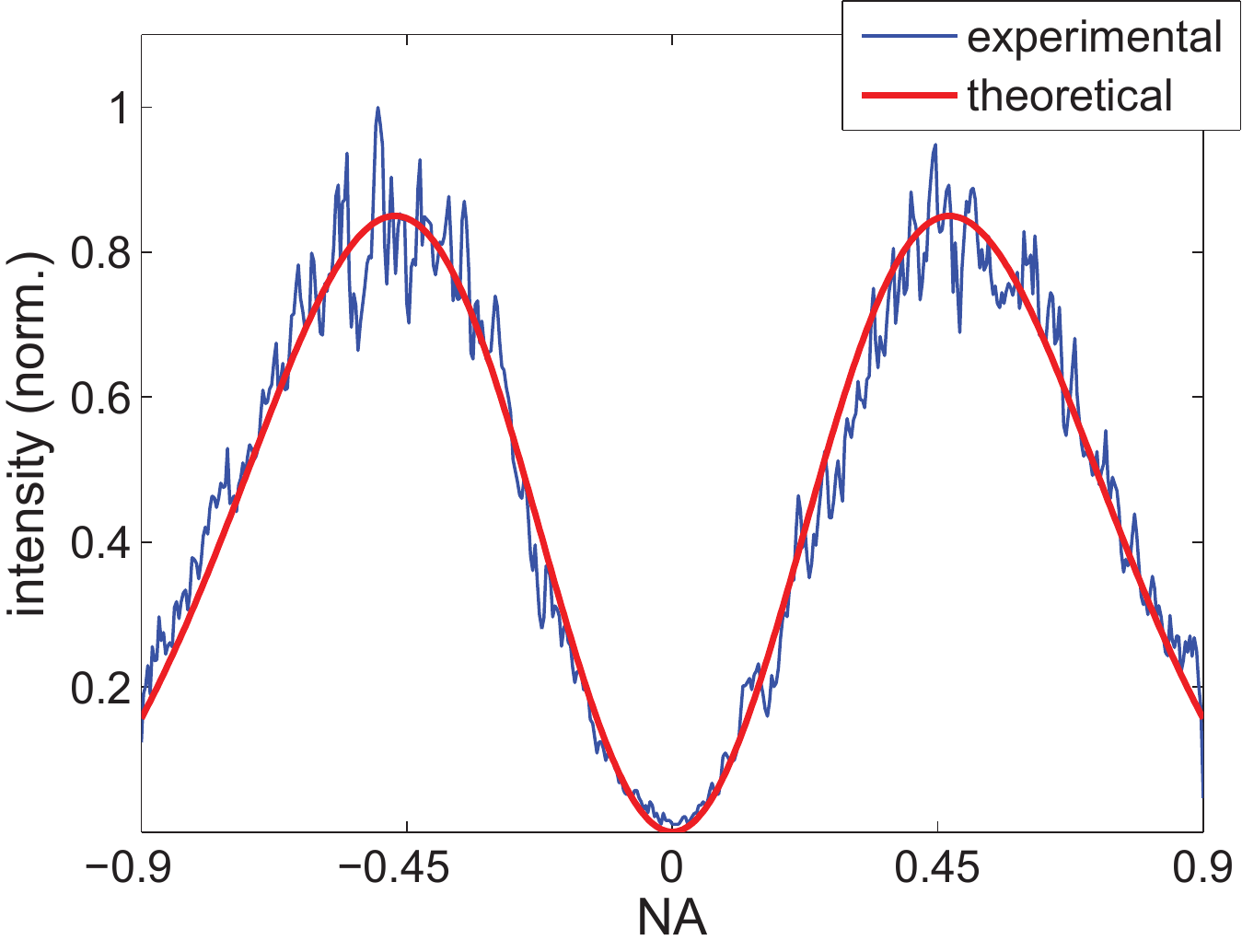}
\caption{Influence of sampling: comparison between the theoretically expected beam cross-section in the back focal plane of the collection objective and the experimentally retrieved cross-section for a radially polarised input beam with $w_0/f = 0.69$, corrected by the sampling-factor $\cos^2 \theta$.}
\label{fig:sampling-BFP}
\end{figure}

To measure the energies transmitted to certain solid angles experimentally, the back focal plane of the microscope objective collecting the transmitted and forward scattered light is imaged onto a CCD-camera. This introduces a sampling effect on the recorded back focal plane images, which has to be corrected when comparing the resulting experimental and theoretical transmitted energies. The sampling results from the finite pixel size of the CCD-camera being related to finite integration angles in the spherical coordinate system of the transmitted multipole components. Thus, the effective integration area of each pixel gets transformed from $\mathrm{d}F= \Delta x \Delta y$ to $\mathrm{d}F'= \cos \theta \Delta \theta \Delta \varphi$ (see Ref. \onlinecite{Lindlein2007_S}). Measuring the electric energy density $\left| \mathbf{E} \right|^2_\text{CCD}$, the energy transmitted at a certain collection angle is then $\left| \mathbf{E} \right|^2 = \cos^2 \theta \left| \mathbf{E} \right|^2_\text{CCD}$.
The sampling effect was additionally checked for a radially polarised input beam of known size by rigorously calculating the resulting fields through the focusing and collection objective and comparing the theoretically expected energy density with a CCD-image of the transmitted beam (see Fig. S \ref{fig:sampling-BFP}).

\subsection{Characterisation of the used nanoprobe}
The precise knowledge of the shape and optical properties of the used nanoprobe govern the accuracy of the proposed reconstruction algorithm. Since the employed gold nanoparticle is fabricated via laser-induced melting, we result in a spherical nanoprobe of \SI{82}{\nano\metre} diameter, measured via SEM. To account for possible changes in the material conformation, we measured the effective refractive index of the nanoprobe via Mie-scattering\cite{Mishchenko2002_S}. Thus, the nanoprobe is embedded in a homogeneous medium via index-matching immersion oil ($n=1.52$) and the focussing objective is exchanged with a second oil-immersion objective with a numerical aperture (NA) of 1.3.
By choosing a linearly polarised Gaussian input beam with a small beam width of $w_0/f=0.35$, the effective NA of the system is reduced to 0.5. Positioning the particle in the focal spot of the system, an approximately homogeneous plane wave excitation is achieved over the whole particle.

The normalised intensity collected in reflection and transmission is then linked to the scattering and absorption cross-section of the particle via 
\begin{equation}
 R \approx A_0 \frac{1}{3} Q_\text{sca} \qquad T \approx 1 - A_0 \cdot \left(Q_\text{abs} + \frac{2}{3} Q_\text{sca}\right),
\end{equation}
 where $A_0$ depends on the polarisability of the particle. The factors $\frac{1}{3}$ and $\frac{2}{3}$ follow from the maximum collection angle of $\theta_\text{max}=60^\circ$, where approximately one third of the scattered intensity of a transverse dipole is collected in forward direction as well as in backward direction.

To retrieve the relative permittivity of the particle, the transmittance and reflectance of the embedded particle are measured for a spectral range of $\lambda = 475-\SI{710}{\nano\metre}$. This spectral data is then used to fit a physical model of the particle's relative permittivity to the classical Mie-scattering solution of $Q_\text{sca}$ and $Q_\text{abs}$. The utilised model is based on an extended Drude model with critical point description\cite{Etchegoin2006_S} for the interband transitions and was shown to give excellent results for fitting the permittivity of planar gold films. 
\begin{figure}[hbt]
\centering
\subfloat{\includegraphics[width=0.6\linewidth]{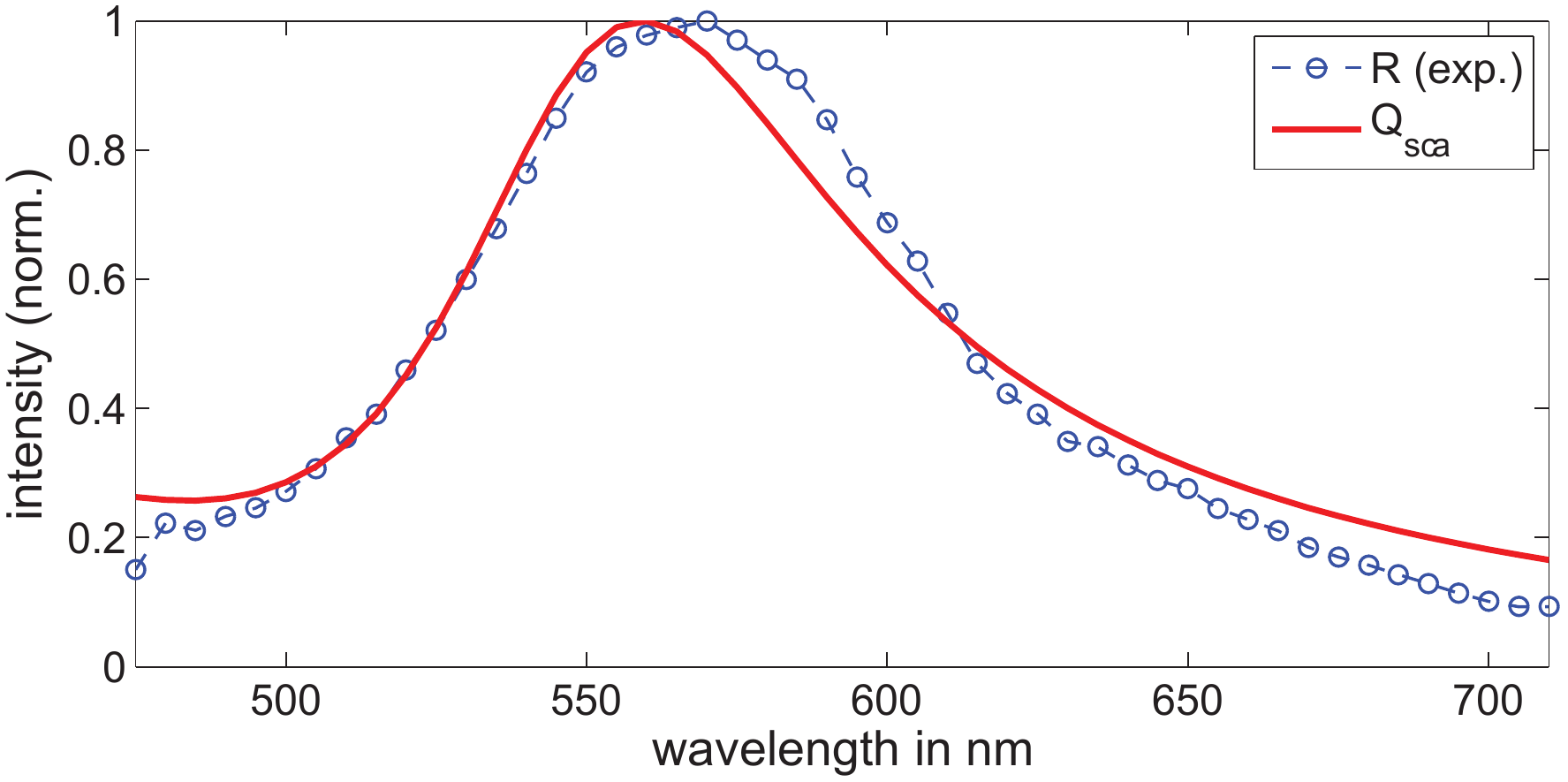}}\\
\subfloat{\includegraphics[width=0.6\linewidth]{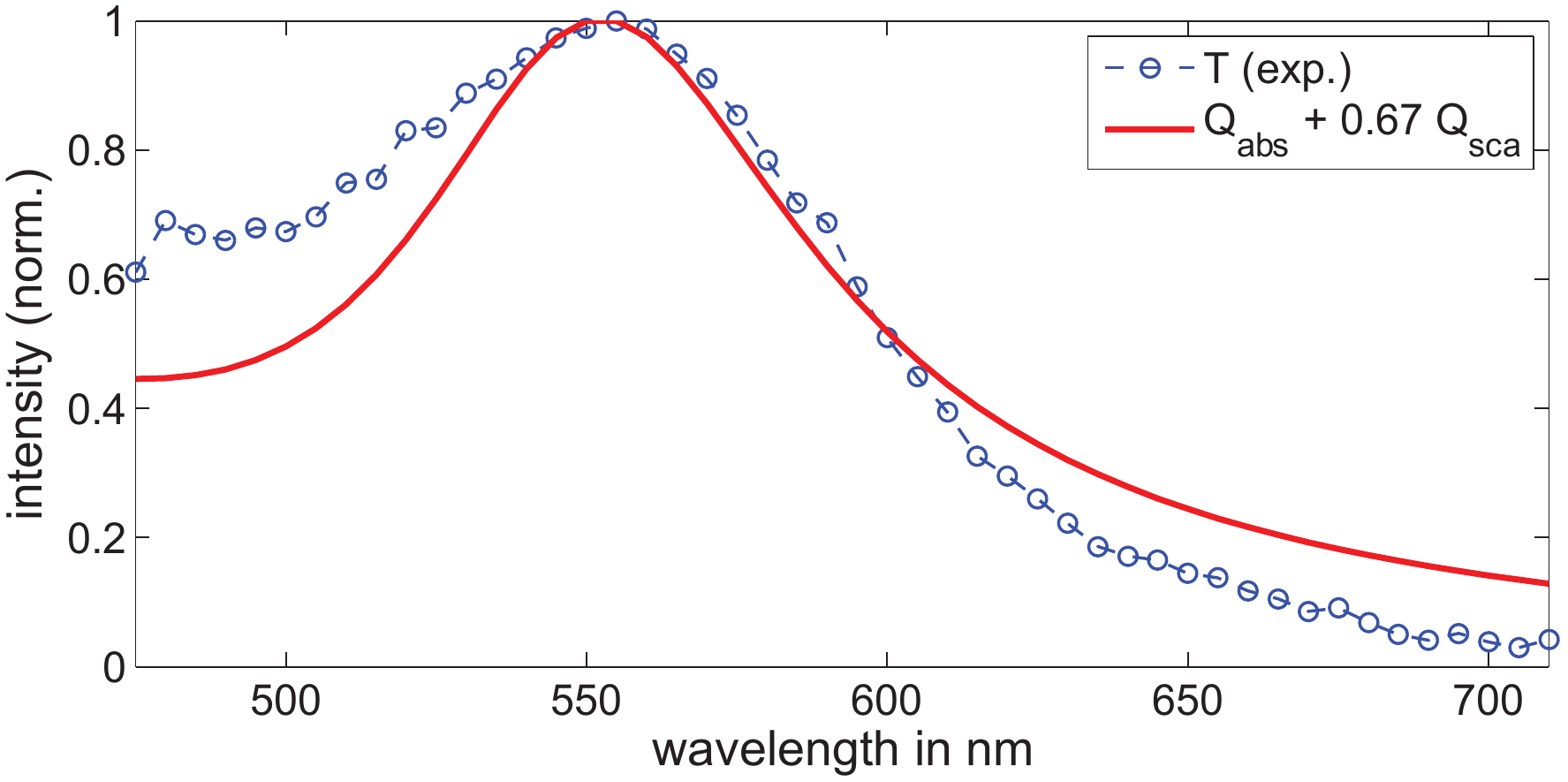}}
\caption{Determination of $\epsilon$: Measured and fitted reflectance (top) and transmittance (bottom) spectrum for the used nanoprobe embedded into a homogeneous medium with $n_m=1.52$.}
\label{fig:resonance}
\end{figure}
The fitted spectral response shows a very good overlap with the measured resonance spectra (see Fig. S \ref{fig:resonance}) and results in $$\epsilon(\lambda=\SI{530}{\nano\metre}) = (-3.0 \pm 0.2) + (2.1 \pm 0.2)\imath$$
for the wavelength used in the reconstruction scheme. The resulting wavelength dependent permittivity additionally reproduces the resonance wavelength of colloidal gold particle solutions with different diameters very well (tested with particles from nanoComposix, Inc.).

\subsection{Effect of scanning and influence of the step size}
When scanning the nanoprobe through the focal field, one has to consider the general effect of translating the probe as well as  the influence of the chose step size. Thus, let us consider one scanning step by a new coordinate frame ($R', \theta ', \varphi '$), related to the old one by $\mathbf{R}=\mathbf{R}'+\mathbf{R_0}$. Here, $\mathbf{R_0}=\left(R_0, \theta _0, \varphi _0 \right)$ is an arbitrary displacement, where the polar and azimuthal coordinate axes remain unchanged. The functions  $\mathbf{M}_{mn}$, $\mathbf{N}_{mn}$ of the old coordinate frame can then be expressed as a sum of the functions  $\mathbf{M}'_{\mu \nu}$, $\mathbf{N}'_{\mu \nu}$ of the new coordinate frame. The corresponding electric field is then expressed as $\mathbf{E}_\text{in}=\sum _{\nu=1}^{\infty}\sum _{\mu=-\nu}^{\nu}  A'_{\mu \nu} \mathbf{N}'_{\mu\nu} + B'_{\mu \nu} \mathbf{M}'_{\mu\nu}$, where the expansion coefficients $A'_{\mu \nu}$ and $B'_{\mu \nu}$ can be calculated from the addition theorem of VSHs\cite{Cruzan1962_S}. In a matrix representation, this corresponds to
\begin{equation}
\left[ \begin{array}{ll}
	\mathbf{B'} \\
	\mathbf{A'}
	\end{array} \right]= \left[ \begin{array}{ll}
	\mathbf{Tr}_{MM}(\mathbf{R}_0) & \mathbf{Tr}_{NM}(\mathbf{R}_0) \\
	\mathbf{Tr}_{MN}(\mathbf{R}_0)  & \mathbf{Tr}_{NN}(\mathbf{R}_0) 
	\end{array} \right] \left[ \begin{array}{ll}
	\mathbf{B} \\
	\mathbf{A}
	\end{array} \right].
\label{eq:add_theor}  
\end{equation}
where $\mathbf{Tr}_{ij}(\mathbf{R}_0)$ are sub-matrices of the translation operator $\mathbf{Tr}(\mathbf{R}_0)$.

To determine an optimized scan step $\Delta \rho < \lambda$ for the experiment, the elements of a translation matrix are usually represented as the sums 
\begin{align}
A_{\mu\nu}^{mn} &= \sum\limits_{p}
a_{mn}^{\mu\nu}(p) j_p(k\Delta \rho) P_p^{m-\mu}(0) \mathrm{e}^{ \mathrm{i}(m-\mu)\varphi_0 }
 \nonumber \\
B_{\mu\nu}^{mn} &= \sum\limits_{p}
b_{mn}^{\mu\nu}(p) j_p(k\Delta \rho) P_p^{m-\mu}(0) \mathrm{e}^{ \mathrm{i}(m-\mu)\varphi_0 },
\label{eq:Amunu}
\end{align}
where $\varphi _0$ denotes the azimuthal coordinate (direction) of the scan step, $|\nu - n| \leq p \leq \nu + n$, $a_{mn}^{\mu\nu}(p)$ and $b_{mn}^{\mu\nu}(p)$ are transition matrix elements\cite{Cruzan1962_S}, $j_i(x)$ denotes the spherical Bessel functions of first kind, and $P_i^j(x)$ stands for the associated Legendre polynomials.
For a small steps $\Delta \rho \approx 0$, the spherical Bessel functions can be approximated to $j_0 \approx 1 + O(\Delta \rho ^2)$, $j_1 \approx k \Delta \rho / 3 + O(\Delta \rho ^3)$, and for $n>2$ to $j_n \approx O(\Delta \rho ^n)$. Translating the nanoprobe by a small distance, the sums in Eq. (\ref{eq:Amunu}) thus contain only two terms: $p=0$ with $m = \mu$ and $p = 1$ with $|m - \mu| = 0, 1$. This implies that only electric dipoles and quadrupoles as well as magnetic dipoles have to be considered. Since the employed nanoprobe responds mainly to electric dipoles, the size of the step has to be chosen, so that a translated magnetic dipole or electric quadrupole results in a measurable change of the electric energy density in the focal plane. This statement can be mathematically formulated as    $\mathbf{Tr}=\hat{\mathbf{1}} + k\Delta \rho \hat {\mathbf{O}}_1$, where $\hat{\mathbf{1}}$ is a diagonal matrix and $k\Delta \rho \hat {\mathbf{O}}_1$ represents small linear changes. The size of the step now has to be chosen, so that the change in the scattering energy $\Delta W_{sca}$ due to the change of the coordinate system is larger than the background noise $W_\text{noise}$ introduced in the experimental measurements.

Our estimation shows, that for our nanoprobe system, a scan step of $\Delta \rho = 25$ nm is large enough to tolerate fluctuations in the energy of up to 8\%, whereas scan steps of $\Delta \rho = 10$ nm are sufficient to tolerate fluctuations of up to 3\%.

\subsection{Additional sources of uncertainties}
Due to the desired sub-wavelength resolvability of the focal field reconstruction, several experimental and numerical uncertainties have to be considered. From an experimental point of view, the exact knowledge of the optical axis of the collection objective as well as its optical properties concerning aberrations and polarisation dependent transmission coefficients have to be exactly known. A mispositioned optical axis corresponds to a rotation of the chosen solid angles and thus introduce a changed weighting of the field components. Pinpointing the axis position in the back focal plane to better than $\text{k}_0/100$ results thus in an accurate reconstruction of the electric field components up to a relative strength of $10^{-3}$.  Any strong aberration of the collection objective on the other hand can not be easily distinguished from aberrations in the reconstructed field and thus lead to ambiguities. We found that for current optimized high-NA oil immersion objectives this error source is negligible.   

Concerning the nanoprobe, its exact optical material parameters and its geometrical shape have to be determined accurately, as mentioned in S.4. A spherical probe is of distinct advantage here, making the size estimation via SEM and the material parameter determination via Mie-scattering routinely feasible. 

\begin{figure}[hbt]
\centering
\includegraphics[width=\linewidth]{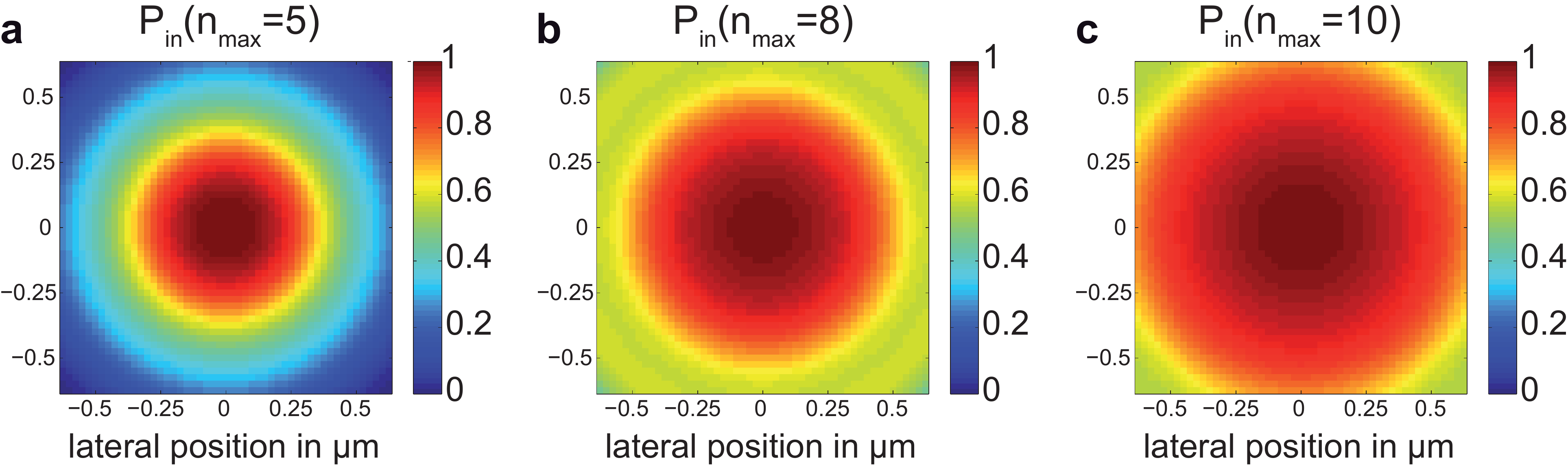}
\caption{Relative power of a radially polarised focal field (with the same parameters as in the main manuscript) contained in the first \textbf{a} $n_\text{max}=5$, \textbf{b} $n_\text{max}=8$ and \textbf{c} $n_\text{max}=10$ multipole orders when being translated through the focal plane. The lateral position corresponds to a relative shift of the beam axis with respect to the expansion axis of the multipoles.}
\label{fig:beamenergy}
\end{figure}

From a theoretical point of view, the maximum multipole order of the focal field sets a limit to the lateral dimension of the focal field as well as the accuracy to reconstruct small structures of the field. Due to the translation of the probe through the focal field, the expansion point of the scattered and extinct fields is shifted with respect to the optical axis. Thus, power is distributed to higher order multipoles, leading to a distorted scattering response when considering only $n$ multipole orders. This effect can be seen in Fig. S \ref{fig:beamenergy}, where the power contained in 5, 8 and 10 multipole orders is shown for a radially polarized focal field with the same parameters as in the main manuscript when the beam is shifted with respect to the central position. While 5 multipole orders still show a significant drop in power at the outer borders of the scan area, 8 multipoles give already a fairly good account of the beam power. With 10 multipole orders, the influence of the lateral shift is then negligible for the dimensions of the actual focal field. Considering tightly focused fields with focal spot diameters smaller than the wavelength, a multipole order of $n_\text{max}=8$ thus shows accurate reconstruction results within the focal spot (see also Refs. \onlinecite{Hoang2012_S,Mojarad2008_S}). Additionally, the choice of the collected solid angles is crucial to achieve unambiguous results as shown in S.2.

\subsection{Reconstruction of further vectorial focal field distributions}
\begin{figure*}[hbt]
\centering
\includegraphics[width=\linewidth]{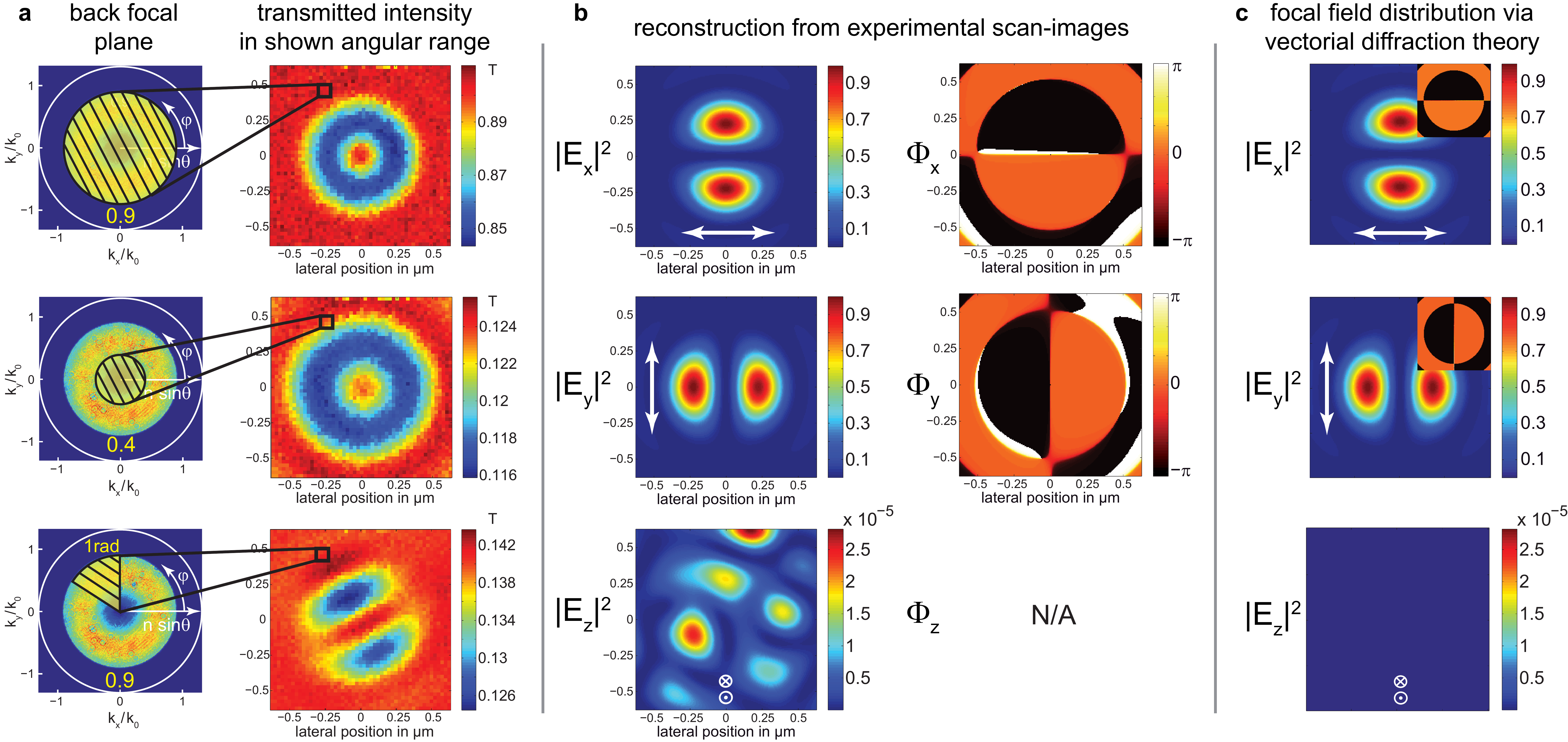}
\caption{Results for an azimuthally polarized vector beam. \textbf{a} Image of the back focal plane of the collection objective with an NA of 1.3 for one position of the nanoprobe relative to the focal field. The measured intensity scan-images in transmission for a wavelength of \SI{530}{\nano\metre} correspond to three different integration solid angles with a full aperture for an NA of 0.9 and 0.4 as well as an azimuth angle of $\varphi_a=1\ \text{rad}$ for an NA of 0.9. The intensity is normalized to the total intensity of the input beam. \textbf{b} Squared electric field components $\left| E_\text{i} \right|^2$ and relative phase $\Phi_\text{i}$ in the focal plane reconstructed from the measured intensity distributions. \textbf{c} Energy density distribution for the electric field components in the focal plane of the same beam calculated via vectorial diffraction integrals. The insets show the calculated phase distribution in the same scale and colormap as in (b).}
\label{fig:results_suppl}
\end{figure*}
To confirm the capabilities of the introduced vectorial reconstruction scheme, we experimentally measured different tightly focused field configurations. Figure S \ref{fig:results_suppl} shows the experimental results of an azimuthally polarised input beam for three different ranges of integration in the back focal plane of the collection objective (Fig. S \ref{fig:results_suppl}(a)) as well as the reconstructed (Fig. S \ref{fig:results_suppl}(b)) and theoretically calculated (Fig. S \ref{fig:results_suppl}(c)) focal field distributions. The reconstructed and numerically evaluated electric field components show an excellent overlap, with only a weak residual longitudinal electric field component stemming from either uncertainties in the numerical reconstruction or noise of the measurement.
From the integrated intensity collected in a solid angle of 1 rad, details of the interferometric information down to a lateral variation of $\SI{100}{\nano\metre}$ or $\lambda/5$ can be resolved.

\end{document}